\documentclass{iau}

\usepackage{amsmath}
\usepackage{graphicx}
\usepackage{multirow}

\begin{document}

\lefttitle{A.\,P.\,Milone}
\righttitle{Multiple Populations in Globular Clusters with JWST}

\jnlPage{1}{7}
\jnlDoiYr{2023}
\doival{10.1017/xxxxx}

\aopheadtitle{Proceedings IAU Symposium No. 377}
\editors{F. Tabatabaei \& B. Barbuy}

\title{Multiple Stellar Populations in Globular Clusters with JWST}

\author{A.\,P.\,Milone$^{1,2}$}
\affiliation{$^{1}$ Dipartimento di Fisica e Astronomia ``Galileo Galilei'', Universita' di Padova, Vicolo dell'Osservatorio 3, Padova, IT-35122}
\affiliation{$^{2}$Istituto Nazionale di Astrofisica - Osservatorio Astronomico di Padova, Vicolo dell'Osservatorio 5, Padova, IT-35122}
\begin{abstract}
  I present the first evidence of multiple populations in the globular cluster (GCs) 47\,Tucanae based on images collected with the near-infrared camera (NIRCam) on board the {\it James Webb Space Telescope} ({\it JWST}\,). While NIRCam photometry is poorly sensitive to multiple populations among stars brighter than the main-sequence (MS) knee, the M-dwarfs more-massive than $\sim$0.1 $\mathcal {M}_{\rm \odot}$ define a wide $m_{\rm F115W}-m_{\rm F322W2}$ color range due to multiple populations. The star-to-star color differences are mostly due to the different amounts of water vapor (hence oxygen) that affect the spectra of M-dwarfs.
   The chromosome map unveils an extended first population (1P) composed of M-dwarfs with different metallicities and three main groups of second-population (2P) stars that are depleted in oxygen with respect to the 1P. I present the discovery of an MS of very-low-mass stars ($\mathcal{M}_{\rm \odot} <0.1 \mathcal{M}_{\rm \odot}$) and tentatively associated it with a sequence composed of O-rich stars alone.
\end{abstract}


\maketitle

\begin{figure}[t]
  \centerline{\vbox to 6pc{\hbox to 10pc{}}}
  \includegraphics[height=9.0cm,trim={0.1cm 0cm 0.4cm 0.0cm},clip]{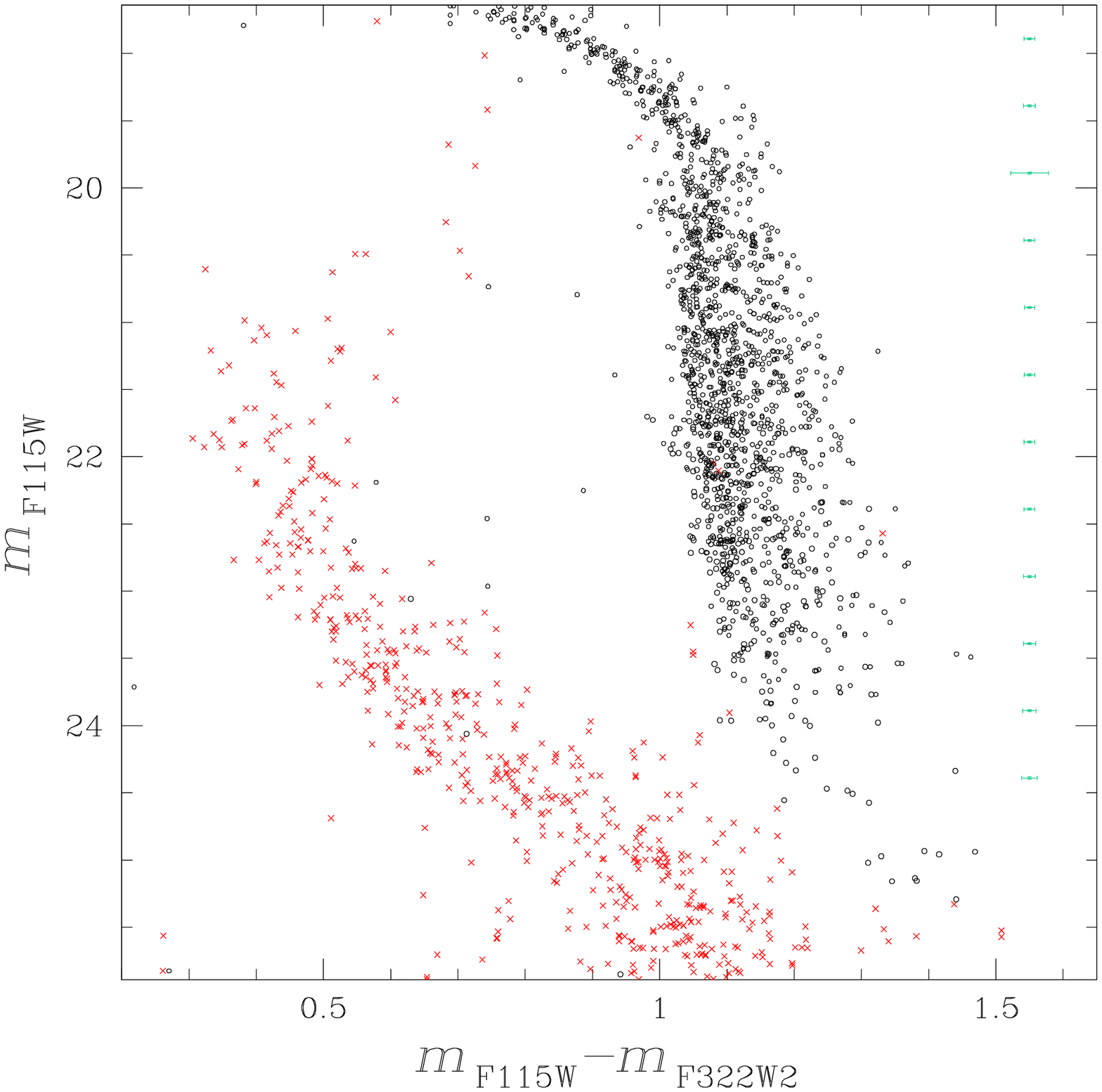}
  \includegraphics[height=8.9cm,trim={0.2cm 0cm 9.5cm 0.0cm},clip]{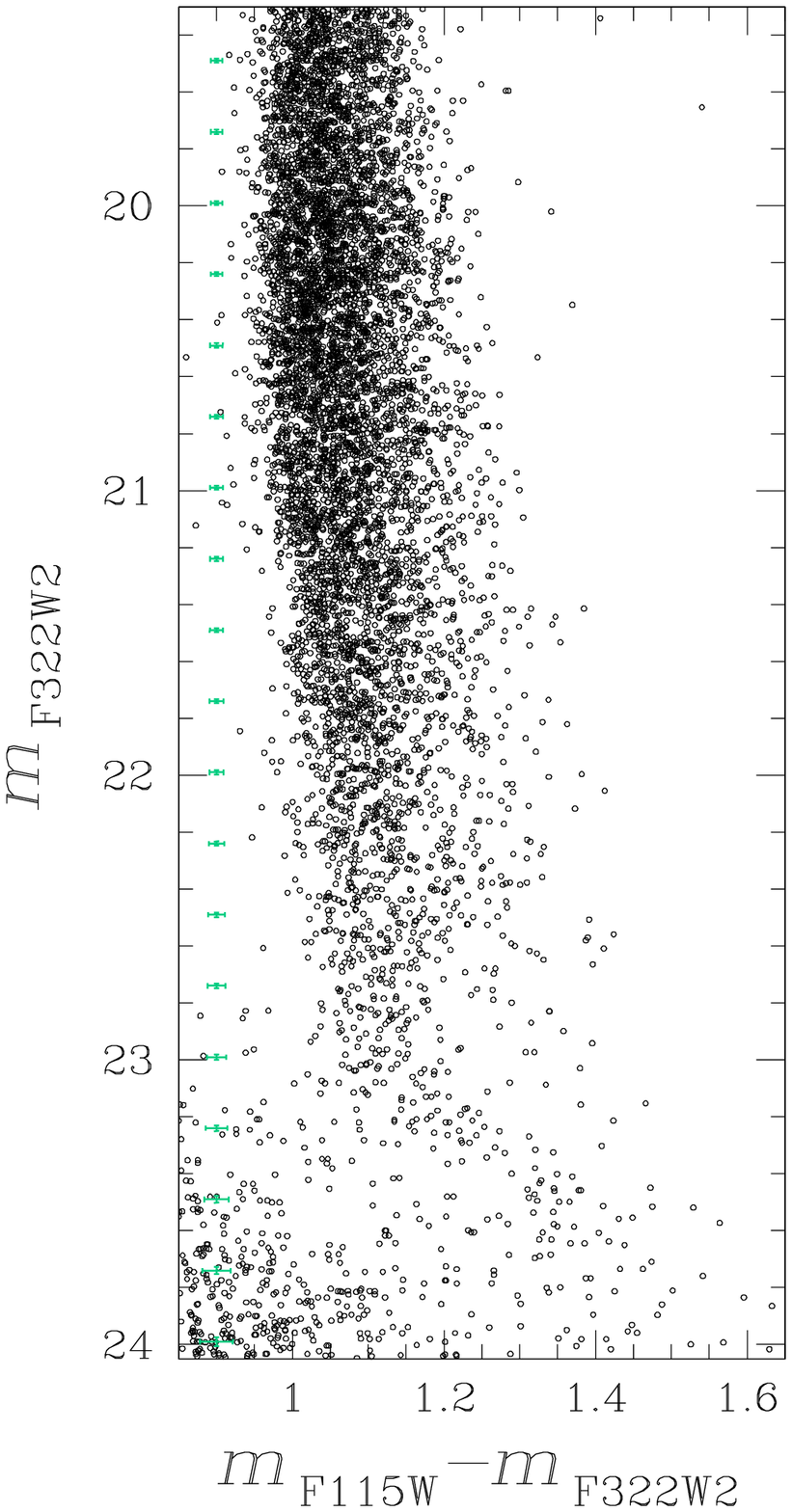}
  \caption{ \textit{Left.} $m_{\rm F115W}$ vs.\,$m_{\rm F115W}-m_{\rm F322W2}$ CMD of stars in the field of view of 47\,Tucanae with available proper motions. The proper-motion selected cluster members are represented with black points, while the red crosses mark field stars.  \textit{Right.} Zoom of the $m_{\rm F322W2}$ vs.\,$m_{\rm F115W}-m_{\rm F322W2}$ CMD for all the stars in the NIRCam fild of view around the MS region below the knee (Milone et al.\,2023). NIRCam photometry is derived from images collected within the program GO-2560 (PI.\,A.\,F.\,Marino).}
  \label{fig:pm}
\end{figure}
\section{Introduction}
Most globular clusters (GCs) host groups of stars with different chemical compositions. A first stellar population (1P) that resembles Milky-Way field stars with the same metallicity and one or more populations (2P) of stars enhanced in helium, nitrogen, sodium, and aluminum and depleted in carbon and oxygen (see reviews by Milone \& Marino\,2022, Bastian \& Lardo 2018, Gratton et al.\,2012). 
Nowadays, the origin of multiple populations in GCs is one of the most-debated open issues of stellar astrophysics. Some scenarios predict that GCs experienced two or more star-formation bursts and that 2P stars formed from polluted gas that is ejected by more-massive 1P stars (e.g.\,Renzini et  al.\,2022, D'Antona et al.\,2016,  Denissenkov \& Hartwick 2014, Decressen et al.\,2007). The main alternative scenarios suggest that all GC stars are coeval and the chemical composition of 2P stars is due to the accretion of polluted material onto pre-main-sequence (MS) stars (e.g.\,Gieles et al.\,2018).

The multiple populations define distinct sequences in photometric diagrams constructed with magnitudes taken in appropriate bands (Milone \& Marino 2022). In particular, the UV filters of the {\it Hubble space telescope} ({\it HST}\,) allowed identifying and characterizing multiple populations along the upper MS, the sub-giant branch (SGB), the red giant branch (RGB), the horizontal branch, and the asymptotic giant branch in a large sample of GCs (e.g.\,Milone et al.\,2017, Lagioia et al.\,2019, Dondoglio et al.\,2021).
However, these studies are focused on stars more massive than  $\sim 0.6 \mathcal{M_{\odot}}$. Indeed it is not possible to derive precise UV photometry of faint stars with present-day facilities.
Photometry in infrared bands allowed for overcoming this limitation. As an example, the F110W and F160W bands of {\it HST} are efficient tools to identify multiple populations among M-dwarfs. The reason is that the F160W filter is significantly affected by absorption from water and other molecules that include oxygen, while F110W photometry is poorly affected by oxygen abundance. Since 2P stars are oxygen-depleted, they have redder $m_{\rm F110W}-m_{\rm F160W}$ colors than 1P stars with similar luminosities (Milone et al.\,2012, 2019, Dotter et al.\,2015, Dondoglio et al.\,2022).
In the following, I summarize the first results on multiple populations with {\it JWST}.

\begin{figure}[t]
  \centerline{\vbox to 6pc{\hbox to 10pc{}}}
  \includegraphics[height=6.55cm,trim={0.0cm 0cm 2.9cm 6.5cm},clip]{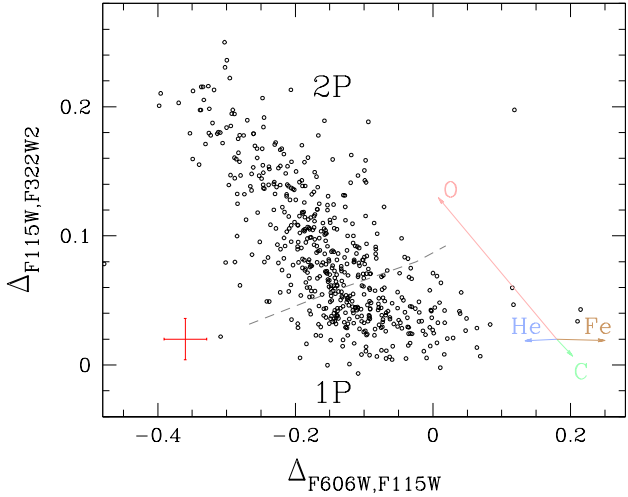}
  \caption{$\Delta_{\rm F115W,F322W2}$ vs.\,$\Delta_{\rm F606W,F115W}$ chromosome map of M-dwarfs in 47\,Tucanae. The dashed line separates the bulk of 1P stars from the 2P stars. The arrows, which for clarity are shifted on the bottom-right corner, indicate the effect of changing the abundances of He, C, N, O, and Fe by $\Delta$Y=0.05, $\Delta$[C/Fe]=$-$0.25 dex, $\Delta$[O/Fe]=$-$0.25 dex, and $\Delta$[Fe/H]=0.1 dex.}
  \label{fig:ChM}
\end{figure}

\section{A NIRCam view of multiple populations in globular clusters}
Based on synthetic spectra and stellar evolution models, Milone et al.\,(2023) and Ziliotto et al.\,(2023) investigated the effect of multiple populations on synthetic diagrams constructed with NIRCam photometry. They simulated metal-rich ([Fe/H]=$-$0.75), metal-intermediate ([Fe/H]=$-$1.5), and metal-poor ([Fe/H]=$-$2.3) isochrones with ages of 13 Gyr that account for the chemical composition of 1P and 2P stars.
NIRCam magnitudes are poorly sensitive to multiple populations among the MS segment brighter than the knee, the SGB, and most RGB, with the exception of stellar populations with extreme helium variations and stars near the RGB tip (see also Salaris et al.\,2019).

 Below the MS knee, 1P and 2P M-dwarfs of all simulated metallicities populate distinct MSs in various CMDs constructed with NIRCam filters. The F090W$-$F300M and F090W$-$F277W colors provide the maximum separations between the MSs. Other CMDs based on the F115W and F322W2 filters and with F150W2 and F322W2 filters are efficient tools to disentangle 1P and 2P M-dwarfs. The MS split is mostly due to the different amounts of water vapor blanketing in the F300M, F277W, and F322W2 filters, as 2P stars have lower O, hence low H$_{2}$O content, compared to 1P stars that are O-, hence H$_{2}$O-rich.
In addition, metal-rich isochrones show large F070W magnitude differences between 1P and 2P M-dwarfs, and a similar conclusion can be extended to the F606W filters of {\it HST}.

\section{NIRCam observations of multiple populations in 47\,Tucanae}
To test the theoretical calculations, I present results, I present results from high-precision photometry obtained from NIRCam images collected as part of (GO-2560, PI A.\,F.\,Marino).
 In the $m_{\rm F115W}$ vs.\,$m_{\rm F115W}-m_{\rm F322W2}$ CMD plotted in the left panel of Figure\,\ref{fig:pm}, stellar proper motions are used to separate the bulk of 47\,Tucanae cluster members (black dots) from the background Small Magellanic Cloud stars (red crosses). 
The CMD of cluster stars reveals that above the knee, the $m_{\rm F115W}-m_{\rm F322W2}$ color width of the MS is similar to the color broadening due to observational errors, thus demonstrating that NIRCam photometry alone is poorly sensitive to multiple stellar populations among bright MS stars.
On the contrary, stars fainter than the MS knee exhibit a wide color broadening, thus unveiling multiple populations among M-dwarfs with masses between $\sim 0.10-0.45 \mathcal{M}_{\odot}$.  The majority of these stars populate the bluest portion of the MS and correspond to O-rich stars ([O/Fe]$\sim$0.4), but a tail of O-poor M-dwarfs exhibit bluer colors thus resulting in an MS that is up to nearly 0.25 mag-wide at $m_{\rm F115W} \sim 23$.

Intriguingly, the MS color width suddenly decreases around $m_{\rm F322W2} \sim 22.8$ mag, where the MS of very-low-mass stars also turns to red colors. Most MS stars in the mass interval from $\sim 0.1 \mathcal{M}_{\odot}$  towards the hydrogen-burning limit define a narrow MS, whose width is comparable with the color broadening due to observational errors alone. This fact could suggest that most very-low-mass stars are O-rich and indicate the lack of O-poor 2P stars among stars with $\mathcal{M}<0.1 \mathcal{M}_{\odot}$ (Milone et al.\,2023).

Multiple populations are also evident in CMDs constructed with both {\it HST} and {\it JWST} photometry. The $m_{\rm F115W}$ vs.\,$m_{\rm F606W}-m_{\rm F115W}$ CMD reveals that most M-dwarfs of 47\,Tucanae are distributed in the middle of the MS, but two less-numerous populations define MSs with bluer and redder colors than the bulk of MS stars.
 The chromosome map (ChM) derived from the  $m_{\rm F115W}$ vs.\,$m_{\rm F606W}-m_{\rm F115W}$ and $m_{\rm F115W}$ vs.\,$m_{\rm F115W}-m_{\rm F322W2}$ CMD is plotted in Figure\,\ref{fig:ChM}. It allows us to identify the 1P stars, which are located around the origin of the ChM, and three main groups of stars, which span a wide pseudo-color interval that ranges from $\Delta_{\rm F115W,F322W2} \sim 0.06$ mag to $\sim$0.25 mag (Milone et al.\,2023).

 The ChM unveils an extended 1P sequence that corresponds to an internal variation of $\Delta$[Fe/H]$\sim$0.12 dex. The majority of 1P M-dwarfs share low iron content, while the remaining stars have higher metallicities, in close analogy with what is observed for RGB stars. This finding provides the first evidence of star-to-star iron variation among M-dwarfs and corroborates the evidence that 1P stars of most GCs have different metallicities (e.g.\,Milone et al.\,2015, Marino et al.\,2019, Legnardi et al.\,2022).

 The color separation between 1P and 2P M-dwarfs is reproduced by isochrones where the oxygen abundances of the distinct populations correspond to the [O/Fe] values inferred from high-resolution spectroscopy of RGB stars.
 The evidence that multiple populations among stars with different masses share similar light-element abundances may help to disentangle the formation scenarios. While multiple-generation scenarios may imply that 2P stars with different masses have the same chemical composition, results on the M-dwarfs 47\,Tucanae may provide a strong constraint for those scenarios where the chemical composition of 2P stars is due to accretion of polluted material onto pre-MS stars

\end{document}